\documentclass[russian,english]{article}
\usepackage[T1]{fontenc}
\usepackage[koi8-r]{inputenc}
\usepackage{amsmath}
\usepackage{amssymb}
\usepackage{esint}

\makeatletter
\@ifundefined{date}{}{\date{}}

\newtheorem{theorem}{Theorem}

\newtheorem{corollary}{Corollary}
\newtheorem{lemma}{Lemma}

\newtheorem{remark}{Remark}

\usepackage{babel}

\makeatother

\usepackage{babel}
\begin{document}

\title{Dynamical phase transition in the simplest molecular chain model }

\author{V. A. Malyshev, S. A. Muzychka}
\maketitle
\begin{abstract}
We consider the dynamics of the simplest chain of large number $N$
of particles. We find, in the double scaling limit, the partition
of the parameter space onto two domains, where for one domain the
supremum (over all time interval $(0,\infty)$ of relative extension
tenda to $1$ as $N\to\infty$. and for the other this supremum tends
to infinity as $N\to\infty$. 
\end{abstract}

\section{Introduction}

For mathematical models of equilibrium statistical physics one needs
stability, that is finiteness of the partition function in the finite
volume. This stability condition provides good approximation for many
phenomena in gases, liquids and even condensed matter. However, for
example, any analysis of the models for expansion or destruction of
condensed matter rests on the problem that the volume is not fixed
and it is necessary to consider finite number of particles in the
infinite volume. For realistic interactions (when the potential disappears
at infinity) such system is not stable that is Gibbs distribution
does not exist. It is common to say that the system is metastable,
see for example \cite{Penrose}. 

For finite number of particles, it is necessary then to prove that
the system does not quit some bounded region of the phase space (does
not dissociate into pieces). As this region depends on all parameters
of the model, for large number of particles it is convenient to use
the method, the so called double (sometimes it is better to say multiple)
scaling limit in physics. In our case all parameters are scaled with
respect to large number $N$ of particles. Then one can get asymptotic
estimates in the large $N$ limit, exhibiting the phase transition
domain of parameters. 

We consider one-dimensional system of $N$ classical point particles
(molecules) of the same mass $m$ and initial configuration at time
$t=0$ 
\begin{equation}
0=z_{0}(0)<z_{1}(0)=a<z_{2}(0)=2a<\ldots<z_{N-1}(0)=(N-1)a\label{initial}
\end{equation}
for some $a>0$. The dynamics of such system is defined by the following
hamiltonian 
\begin{equation}
H=\sum_{k=1}^{N-1}\frac{p_{k}^{2}}{2m}+\sum_{k=1}^{N-1}V(z_{k}-z_{k-1})-fz_{N-1}\label{hamiltonian}
\end{equation}
It is assumed that one of the particles $z_{0}$ stays permanently
at zero, and $z_{N-1}$ is subjected to the constant extarnal force
$f>0$. Concerning the function $V(z)$ it is assumed that $V(z)\to\infty$
as $z\to0$, $V(z)\to0$ as $z\to\infty$, $V(z)$ is convex on the
interval $(0,b)$, concave on $(b,\infty)$ and has the unique minimum
$V(a)<0$ at some $a>0$, where $a<b<\infty$. Of course, Gibbs distribution
with such hamiltonian does not exist, and two approaches are possible:
\begin{enumerate}
\item to change $V(z)$ for large $z$ so that $V(z)\to\infty$ as $x\to\infty$,
then one can use Gibbs ideology to study thermal and elastic extension
in equilibrium. See the book \cite{Kardar} where there are many related
applied problems, and, in particular, by simple calculation for harmonic
chain, the growth of variance of extension, linear in temperature,
had been demonstrated. However, as it was shown earlier in \cite{Mal},
the mean extension can be linear in $T$ only for non-harmonic case.
\item to find a neighborhood $O(a)$ of the minimum such that, for initial
data (\ref{initial}) the trajectory stays in $O(a)$ forever. 
\end{enumerate}
Here we follow the second approach, assuming additionally that in
the vicinity of point $a$ the potential has the quadratic form
\begin{equation}
V(z)=\frac{\kappa}{2}(z-a)^{2}\label{quadratic}
\end{equation}
Now we formulate this more exactly. It is not difficult to calculate
the extension of such chain in the static situation (at zero temperature),
that is to find the unique fixed point (minimum of $H$), see below.
However it is logical to study the dynamics of this particle system
and get good estimates for the functional
\begin{equation}
A=A(N,l,f,\kappa,m)=\max_{1\leq k,k+l\leq N-1}\sup_{t\in(0,\infty)}|z_{k+l}(t)-z_{k}(t)|\label{sup}
\end{equation}
for large $N$ and various $l>0$. In despite of evident simplicity
of the model, the main result of the paper - estimates for the maximal
(over all time interval) deviations from the initial crystal structure
is nontrivial and uses some facts from number theory. The question
is that, although the model has a simple fixed point, but as the model
is hamiltonian, then there is no any convergence to this fixed point.
Thus, the problem is to estimate how far the trajectories can be from
this point. We start with fixed number $N$ of particles and find
the neighborhood which the trajectory never leaves. Then in the limit
$N\to\infty$. we find the phase transition between the scalings of
the parameters, for which the crystal structure changes only slightly
on all time interval, and the scalings for which this supremum grows
with $N$. 

We want to note that there are many papers concerning other problems
for one-dimansional modelss. Most popular are the Fermi-Pasta-Ulam
models \cite{Gallavotti} and the Frenkel-Kontorova model \cite{Braun}.
One should also mention the papers \foreignlanguage{russian}{\cite{braides,Braun,braun_schmidt,E_Ming_1,E_Wing_2},
erasing from the book \cite{Born_Huang}, where multi-dimensional
static models were considered with the goal to derive equations of
linear elasticity from micromodels. }

\section{Main result}

Defining the deviations $x_{k}(t)=z_{k}(t)-ka,k=0,...,N-1$, we have
the following hamiltonian system of linear equations 
\begin{equation}
\begin{cases}
\ddot{x}{}_{0}(t) & =0,\\
\ddot{x}{}_{k}(t) & =\omega_{0}^{2}(x_{k-1}-2x_{k}+x_{k+1}),\\
\ddot{x}_{N-1}(t) & =\omega_{0}^{2}(-x_{N-2}+x_{N-1})+f_{0},
\end{cases}k=1,\ldots,N-2\label{eq:main}
\end{equation}
with initial data (\ref{initial}) and $v_{k}(0)=\dot{x}_{k}(0)=0,k=1,...,N-1$.
Here we denoted the proper frequency $\omega_{0}^{2}=\kappa/m$, and
$f_{0}=f/m$.

Introduce the following auxiliary function

\begin{equation}
F_{N}(x)=x\ln\frac{N}{x},x>0,\label{eq:F}
\end{equation}
and our main parameter $\sigma=\frac{f}{\kappa}=f_{0}/\omega_{0}^{2}$.
Let is agree that the constants denoted further by $c,c_{i},const$,
do not depend on $N,l,f_{0},\omega_{0}$ and $a.$ The main estimate
is as follows.

\begin{theorem}Fix some $\varepsilon\in\left(0,1\right)$, then for
any $k,l\in\mathbb{N},$ such that 
\begin{equation}
0\leq k<k+l\leq(1-\varepsilon)N,\label{eq:condition2}
\end{equation}
the following inequalities hold 
\begin{align}
\sigma(l+c_{1}F_{N}(l))\leq & \sup_{t\geq0}\left(x_{k+l}(t)-x_{k}(t)\right)\leq\sigma(l+c_{2}F_{N}(l))\label{bound_main-1}\\
\sigma(l-c_{3}F_{N}(l))\leq & \inf_{t\geq0}\left(x_{k+l}(t)-x_{k}(t)\right)\leq\sigma(l-c_{4}F_{N}(l))\label{bound_main-2}
\end{align}
for some $c_{1},c_{2},c_{3},c_{4}>0$, where $c_{1},c_{3}$ may depend
on $\epsilon$.

\end{theorem}

Further on we use the procedure, called in physics <<double scaling
limit>>. Namely, we put $a=\frac{1}{N}$ and will consider various
scalings of $\sigma=\sigma(N)$.

We use the following notation: for positive functions $f(x)\simeq g(x),x\in\Lambda$,
for some domain $\Lambda$, if there exist such $c_{1},c_{2}>0,$
that on all domain of definition $c_{1}g(x)\leq f(x)\leq c_{2}g(x).$ 

\begin{corollary}

Under the conditions of theorem 1
\[
|x_{k+l}(t)-x_{k}(t)|\simeq\sigma(N)l\ln\frac{N}{l}
\]

\end{corollary}

As the characteristics of this phase transition we use the maximal
relative extension (for $l=1$) 
\[
\frac{A}{a}=NA
\]
under the strength $f$. We have $a^{-1}A\to1$, if $\sigma(N)N\ln N\to0$
as $N\to\infty$, and $a^{-1}A\to\infty$, if $\sigma(N)N\ln N\to\infty$.
More exactly, if $\sigma<\frac{c}{N\ln N}$ for sufficiently small
$c>0$, then the distances will never leave some neighborhood $(\frac{1-\epsilon}{N},\frac{1+\epsilon}{N})$
of $\frac{1}{N}$.

\paragraph{Comparison with equilibrium phase transition}

For quadratic hamiltonian (\ref{quadratic}) a fixed point always
exists and is unique, and moreover for any $k$ the distances $z_{k}-z_{k-1}=h=a+\sigma$.
That is why the static phase transition is as follows: $\frac{h}{a}\to\infty$,
if $\sigma N\to\infty$ and $\frac{h}{a}\to1$, if $\sigma N\to0$.
Thus static and dynamic phase transition differ only by logarithmic
factor. 

The similar fact takes place for more general interactions. Namely,
usually the interaction is assumed to be
\begin{equation}
V(r)=-\frac{c_{n}}{r^{n}}+\frac{c_{m}}{r^{m}}\label{L-D}
\end{equation}
Note that for arbitrary $0<n<m,c_{n}>0,c_{m}>0$ the function $V(r)$
satisfies all properties, formulated above, and moreover the following
holds. If $\max_{a<h\leq b}\frac{dV(h)}{dh}\geq f$ then the hamiltonian
(\ref{hamiltonian}) has the unique minimum, for which all $b\geq z_{k}-z_{k-1}=h>a>0$,
and the value $h$ is defined from the equation 
\begin{equation}
\frac{dV(h)}{dh}=f\label{fixed_point}
\end{equation}
But if $\max_{a<h\leq b}\frac{dV(h)}{dh}<f$, then fixed points do
not exist and, under the action of the force $f$ the chain falls
apart. The following statement concerns the static phase transition
for the interaction (\ref{L-D}). 

\begin{lemma}

If $a=\frac{1}{N}$, then a fixed point exists iff 
\[
\sigma=\frac{f}{\kappa}\leq\frac{1}{N}C
\]
where
\[
\kappa=V''(a),\,\,\, C=C(n,m)=\frac{1}{m-n}((\frac{m+1}{n+1})^{-\frac{n-1}{m-n}}-(\frac{m+1}{n+1})^{-\frac{m-1}{m-n}})
\]
 \end{lemma}

\begin{remark}

For the existence of the Gibbs distribution it is necessary that $V(x)\to\infty$
as $x\to\infty$, Then one can say that for non-zero temperature the
existence of the thermal expansion depends on the third term of the
expansion of $V$ at the point $a$, see\cite{Mal}.

\end{remark}

\section{Proofs}

\subsection{Auxiliary results}

In this section we prove some auxiliary results necessary for the
proof of the main theorem.

\begin{lemma}

The system ($\ref{eq:main}$) has the solution 
\begin{align}
x_{n}(t)= & \sigma\left[n-\frac{1}{(2N-1)}\sum_{m=1}^{2N-2}\gamma_{m,N,n}\cos\omega_{m}t\right],\label{eq:solution}
\end{align}
where 
\begin{equation}
\gamma_{m,N,n}=\frac{1}{\sin^{2}\frac{\pi m}{8N-4}}\sin\frac{\pi m}{2}\cos\frac{\pi m}{4N-2}\sin\frac{\pi nm}{2N-1},\ \omega_{m}=2\omega_{0}\sin\frac{\pi m}{8N-4}.\label{eq:solution2}
\end{equation}

\end{lemma}

Proof. Consider the following auxiliary system of $4N-2$ equations
on the circle (all indices are modulo $4N-2$): 
\begin{equation}
\ddot{y}_{n}=\omega_{0}^{2}(y_{n-1}-2y_{n}+y_{n+1})+f_{0}(\delta_{n,N-1}+\delta_{n,N}-\delta_{n,3N-2}-\delta_{n,3N-1}),\ n=0,\ldots,4N-3\label{eq:Fourier}
\end{equation}
with zero initial conditions. Then we claim that for any $t$ 
\[
x_{n}(t)\equiv y_{n}(t),\ n=0,\ldots,N-1.
\]
In fact firstly, from symmetry of the equations it follows that 
\[
y_{n}=y_{2N-1-n}=-y_{-n}=-y_{2N-1+n},\ n=0,\ldots,N-1.
\]
Thus, $\ddot{y}_{0}=\omega_{0}^{2}(y_{-1}-2y_{0}+y_{1})=-2\omega_{0}^{2}y_{0},$
and, taking the initial conditions into account, it follows that $y_{0}(t)=x_{0}(t)\equiv0.$
Then for $n=1,\ldots,N-2$ 
\[
\ddot{y}_{n}=\omega_{0}^{2}(y_{n-1}-2y_{n}+y_{n+1}).
\]
And finally for $n=N-1:$ 
\[
\ddot{y}_{N-1}=\omega_{0}^{2}(y_{N-2}-2y_{N-1}+y_{N})+f_{0}=\omega_{0}^{2}(y_{N-2}-y_{N-1})+f_{0}.
\]
Thus, the solutions of equations for $x_{n}$ and $y_{n}$ completely
coincide, and we have the result.

Then the system of equations ($\ref{eq:Fourier}$) is easily solved
using Fourier transform 
\begin{equation}
x_{n}=y_{n}=\frac{1}{4N-2}\sum_{m=0}^{4N-3}\alpha_{m}e^{-2\pi inm/(4N-2)},\label{eq:forwardFourier}
\end{equation}
where 
\[
\alpha_{m}=\sum_{n=0}^{4N-3}y_{n}e^{2\pi inm/(4N-2)}
\]
is the solution of the following system of ODE with zero initial conditions
\begin{align}
\ddot{\alpha}_{m}+\omega_{m}^{2}\alpha_{m} & =f_{0}\left(e^{\frac{2\pi im(N-1)}{4N-2}}+e^{\frac{2\pi imN}{4N-2}}-e^{\frac{2\pi im(-N+1)}{4N-2}}-e^{-\frac{2\pi imN}{4N-2}}\right)=\label{eq:alphaEq}\\
 & =f_{0}\left(e^{\frac{\pi im}{2}}-e^{-\frac{\pi im}{2}}\right)\left(e^{\frac{\pi im}{4N-2}}+e^{-\frac{\pi im}{4N-2}}\right)=4if_{0}\cos\frac{\pi m}{4N-2}\sin\frac{\pi m}{2},\nonumber 
\end{align}
It follows 
\[
\alpha_{m}(t)=\frac{4if_{0}}{\omega_{m}^{2}}\cos\frac{\pi m}{4N-2}\sin\frac{\pi m}{2}(1-\cos\omega_{m}t)=\frac{if_{0}}{\omega_{0}^{2}}\cdot\frac{1}{\sin^{2}\frac{\pi m}{8N-4}}\cos\frac{\pi m}{4N-2}\sin\frac{\pi m}{2}(1-\cos\omega_{m}t).
\]
As $y_{n}=-y_{n},$ then $\alpha_{m}=-\alpha_{-m},$ and thus 
\begin{equation}
x_{n}=y_{n}=\frac{1}{4N-2}\sum_{m=0}^{2N-2}\alpha_{m}\left(e^{-2\pi inm/(4N-2)}-e^{2\pi inm/(4N-2)}\right)=-\frac{2i}{4N-2}\sum_{m=1}^{2N-2}\alpha_{m}\sin\frac{\pi nm}{2N-1}.\label{eq:forwardFourier1}
\end{equation}
Finally we have the following formula 
\[
y_{n}=\sigma\frac{1}{(2N-1)}\sum_{m=1}^{2N-2}\gamma_{m,N,n}(1-\cos\omega_{m}t).
\]
To finish the proof we have to verify that 
\[
\frac{1}{(2N-1)}\sum_{m=1}^{2N-2}\gamma_{m,N,n}=n.
\]
Consider the same system ($\ref{eq:Fourier}$), but with initial conditions
\[
y_{n}=y_{2N-1-n}=n\sigma,\ n=-N+1,\ldots,N-1.
\]
It is easy to check that then the system is in the equilibrium $y_{n}(t)\equiv y_{n}(0).$
As a corollary, $\alpha_{m}(t)$ also does not change with time. From
($\ref{eq:alphaEq}$) it follows that the latter is possible only
if 
\[
\alpha_{m}(0)=\frac{4if_{0}}{\omega_{m}^{2}}\cos\frac{\pi m}{4N-2}\sin\frac{\pi m}{2}=i\sigma\frac{1}{\sin^{2}\frac{\pi m}{8N-4}}\cos\frac{\pi m}{4N-2}\sin\frac{\pi m}{2},
\]
and then, from ($\ref{eq:forwardFourier1}$), we have
\[
n\sigma=x_{n}(0)=-\frac{2i}{4N-2}\sum_{m=1}^{2N-2}\alpha_{m}(0)\sin\frac{\pi nm}{2N-1}=\sigma\frac{1}{2N-1}\sum_{m=1}^{2N-2}\gamma_{m,N,n},
\]
and get the desired statement.

Direct substitution of ($\ref{eq:solution}$) shows that we have the
following identity
\begin{equation}
I_{N,k,l}(t):=\sigma^{-1}\left(x_{k+l}(t)-x_{k}(t)-\sigma\cdot l\right)=-\frac{2}{(2N-1)}\sum_{m=1}^{2N-2}a_{m}b_{m}\cos\omega_{m}t\label{difference}
\end{equation}
where
\begin{equation}
a_{m}=\frac{1}{\sin^{2}\frac{\pi m}{8N-4}}\sin\frac{\pi m}{2}\cos\frac{\pi m}{4N-2}\sin\frac{\pi ml}{4N-2},\quad b_{m}=\cos\frac{\pi m(k+l/2)}{2N-1}\label{a_m,b_m}
\end{equation}

We will need also the following facts concerning $a_{m}$ and $b_{m}$. 

\begin{lemma}\label{5_inequalities} The following assertions hold:
\begin{enumerate}
\item $|b_{m}|\leq1$ for any $m\in\mathbb{Z}$; 
\item $a_{m}=0$ for any even $m$; 
\item $|a_{m}|\leq\mathrm{const}\frac{N^{2}}{m^{2}}$ for all $1\leq m\leq2N$; 
\item $\left|a_{m}\right|\simeq\frac{Nl}{m}$ for any odd $1\leq m\leq N/l$;
\item $\frac{\left|b_{m}\right|}{m}+\frac{\left|b_{m+2}\right|}{m+2}\geq\frac{\mathrm{const}}{m}$
for all $m>0$.
\end{enumerate}
\end{lemma}

Proof. The first and second assertions are evident. The third follows
from
\[
\left|a_{m}\right|\leq\sin^{-2}\frac{\pi m}{8N-4}\leq\mathrm{const}\cdot\frac{N^{2}}{m^{2}},
\]
and the fourth follows from the fact that for $1\leq m\leq N/l$
\[
\sin\frac{\pi m}{8N-4}\simeq\frac{m}{N},\ \cos\frac{\pi m}{4N-2}\simeq1,\ \sin\frac{\pi ml}{4N-2}\simeq\frac{ml}{N}
\]
To prove the assertion 5 we should check that there exists such $c>0,$
that for any $m\in\mathbb{Z}$ at least one of the numbers $|b_{m}|,\ |b_{m+2}|$
is not less than $c.$ But from condition ($\ref{eq:condition2}$)
it follows that\foreignlanguage{russian}{
\[
2k+l\leq2(k+l)\leq2(1-\varepsilon)N\Rightarrow0\leq\frac{\pi(2k+l)}{2N-1}\leq\pi(1-\varepsilon),
\]
}and hence, the function 
\[
g(x)=\min\left(|\cos x|,\left|\cos\left(x+\frac{\pi(2k+l)}{2N-1}\right)\right|\right)
\]
is not zero for all $x\in\mathbb{R},$ then from periodicity of $g(x)$
it follows that $\inf_{x}g(x)>0$. But
\[
\min(|b_{m}|,|b_{m+2}|)=g(\pi m(k+l/2)/(2N-1))\geq\inf_{x}g(x),
\]
 and assertion 5 follows. Finally we have for any $m>0$ 
\[
\frac{\left|b_{m}\right|}{m}+\frac{\left|b_{m+2}\right|}{m+2}\geq\frac{\inf_{x}g(x)}{m+2}\geq\frac{\mathrm{const}}{m}.
\]

\begin{lemma}\label{Irrationality}

There exists such $c>0,$ that for any $N\in\mathbb{N}$ and 
\begin{equation}
M=M(N,c)=\left\lfloor cN/\ln\ln N\right\rfloor \label{eq:M(N)}
\end{equation}
the numbers 
\[
\omega_{1}/2\omega_{0},\omega_{2}/2\omega_{0},\ldots,\omega_{M}/2\omega_{0}
\]
are rationally independent (that is there are integers $a_{0},a_{1},\ldots,a_{M}\in\mathbb{Z},$
that $\sum_{m=1}^{M}a_{m}\left(\omega_{m}/2\omega_{0}\right)=a_{0}$).

\end{lemma}

Proof. In this proof we need some facts from the number theory.
\begin{enumerate}
\item (\cite{Niven}, chapter 3) The algebraic degree of the number $e^{2\pi i/n},n\in\mathbb{N},$
equals $\varphi(n),$ where $\varphi(\cdot)$ is the Euler function.
\item (\cite{Hardy}, theorem 328) The following asymptotics holds 
\[
\liminf_{n\to\infty}\frac{\varphi(n)\ln\ln n}{n}=e^{-\gamma},
\]
where $\gamma$ is the Euler-Mascheroni constant. 
\end{enumerate}
For the proof assume the contrary. Putting $z=e^{\frac{\pi i}{8N-4}},$
the rational independence condition can be rewritten as follows 
\[
\frac{1}{2i}\sum_{m=1}^{M}a_{m}(z^{m}-z^{-m})=a_{0}\Rightarrow\left(\sum_{m=1}^{M}a_{m}(z^{m}-z^{-m})\right)^{2}+4a_{0}^{2}=0\Rightarrow\left(\sum_{m=1}^{M}a_{m}(z^{M+m}-z^{M-m})\right)^{2}+4a_{0}^{2}z^{2M}=0.
\]
As in the left-hand side of this equality we have the polynomial of
degree not greater than $4M,$ then the algebraic degree of the number
$z$ does not exceed $4cN/\ln\ln N.$ At the same time from these
two number theoretical facts it follows that there exists $c'>0,$
such that the algebraic degree of $z$ is not less than $c'N/\ln\ln N.$
Then, as $c$ is arbitrary, we get the contradiction.

\begin{corollary}\label{cor_Tor}Under the conditions of lemma $\ref{Irrationality}$
for any $a_{1},\ldots,a_{M}\in\mathbb{R}$ 
\[
\sup_{t\geq0}\sum_{m=1}^{M}a_{m}\cos\omega_{m}t=-\inf_{t\geq0}\sum_{m=1}^{M}a_{m}\cos\omega_{m}t=\sum_{m=1}^{M}|a_{m}|.
\]

\end{corollary}

Proof. From lemma \ref{Irrationality} it follows that the trajectory
$(\omega_{1}t,\omega_{2}t,\ldots,\omega_{M}t)$ is everywhere dense
on the correponding $M$-dimensional torus $T$. From this we have
the desired assertion.

\subsection{Proof of the theorem}

We will check only ($\ref{bound_main-1}$). The formula ($\ref{bound_main-2}$)
can be checked similarly.

\paragraph{Lower bound }

We subdivide the sum in ($\ref{difference}$) into two parts 
\[
I_{N,k,l}(t)=I_{N,k,l}^{1}(t)+I_{N,k,l}^{2}(t):=\sum_{m\leq M(N)}+\sum_{m>M(N)}
\]
and estimate them separately.

\textbf{1)} From the corollary $\ref{cor_Tor}$ it follows that \foreignlanguage{russian}{
\[
\sup_{t\geq0}I_{N,k,l}^{1}(t)=\frac{2}{(2N-1)}\sum_{m=1,m-\text{нечет.}}^{M(N)}\left|a_{m}b_{m}\right|\geq\frac{2}{(2N-1)}\sum_{m=1,m-\text{нечет.}}^{M(N)\wedge[N/l]}\left|a_{m}b_{m}\right|\geq\mathrm{const}\cdot l\cdot\sum_{m=1,m-\text{нечет.}}^{M(N)\wedge[N/l]}\frac{\left|b_{m}\right|}{m}=
\]
\[
=\mathrm{const}\cdot l\cdot\sum_{m=1}^{[(M(N)\wedge[N/l])/4]}\left(\frac{\left|b_{4m+1}\right|}{4m+1}+\frac{\left|b_{4m+3}\right|}{4m+3}\right)\geq\mathrm{const}\cdot l\cdot\sum_{m=1}^{[(M(N)\wedge[N/l])/4]}\frac{1}{m}\geq
\]
\[
\geq\mathrm{const}\cdot l\ln\left(\frac{1}{4}(M(N)\wedge[N/l])\right)\geq\mathrm{const}\cdot l\ln\left(\mathrm{const}\cdot\frac{N}{l\vee\ln\ln N}\right)\geq\mathrm{const}\cdot F_{N}(l)
\]
}where $[.]$ is the integer part and $\wedge$ is the minimum). Here
in the first inequality, we discarded some terms in the sum. In the
second one we used assertion 4. In the third one we used assertion
5 of the lemma \ref{5_inequalities}. In the fourth we used the fact
that 
\begin{equation}
\sum_{m=1}^{n}\frac{1}{m}\simeq\ln n,n\in\mathbb{Z}\label{eq:fact1}
\end{equation}
The inequality 5 follows from the definition of $M(N)$ ($\ref{eq:M(N)}$).
The sixth one can be obtained by separation of two cases: for $l\geq\ln\ln N$
the left-hand part exactly equals the right-hand part, and for $l<\ln\ln N$
we have
\[
l\ln\left(\mathrm{const}\cdot\frac{N}{\ln\ln N}\right)=l\cdot\left(\mathrm{const}+\ln N-\ln\ln\ln N\right)\simeq l\ln N\geq F_{N}(l)
\]

\textbf{2)} We have
\[
\sup_{t\geq0}|I_{N}^{2}(k,l)|\leq\mathrm{const}\cdot N\cdot\sum_{m=M(N)+1}^{2N-1}\frac{1}{m^{2}}\leq\mathrm{const}\cdot N\int_{M(N)}^{\infty}\frac{\mathrm{d}x}{x^{2}}=\mathrm{const}\cdot\frac{N}{M(N)}\leq\mathrm{const}\cdot\ln\ln N
\]
Here in the first inequality we used the statement 3 of lemma 3. The
second one follows from the well-known fact that for $n\to\infty$
\begin{equation}
\sum_{m=n}^{\infty}\frac{1}{m^{2}}\sim\int_{n}^{\infty}\frac{\mathrm{d}x}{x^{2}}\label{eq:fact2}
\end{equation}
In the last inequality we substituted the definition of the function
$M(N)$ ($\ref{eq:M(N)}$).

Joining two cases together we have 
\[
\sup_{t\geq0}I_{N,k,l}(t)\geq\sup_{t\geq0}I_{N,k,l}^{1}(t)-\sup_{t\geq0}|I_{N,k,l}^{2}(t)|\geq\mathrm{const}\cdot F_{N}(l)-\mathrm{const}\cdot\ln\ln N\geq\mathrm{const}\cdot F_{N}(l).
\]

\paragraph{Upper bound }

Now we subdivide the sum ($\ref{difference}$) differently 
\[
I_{N,k,l}(t)=J_{N,k,l}^{1}(t)+J_{N,k,l}^{2}(t):=\sum_{m\leq N/l}+\sum_{m>N/l}
\]
and again estimate them separately.

\textbf{1)} We have
\[
\sup_{t\geq0}|J_{N,k,l}^{1}(t)|\leq\frac{\mathrm{const}}{N}\sum_{m\leq N/l}\frac{Nl}{m}=\mathrm{const}\cdot l\sum_{m\leq N/l}\frac{1}{m}\leq\mathrm{const}\cdot l\ln\frac{N}{l}=\mathrm{const}\cdot F_{N}(l)
\]
Here in the first inequality we used assertions 2 and 4 of lemma 3,
and in the third we used ($\ref{eq:fact1}$).

\textbf{2)} We have\textbf{ 
\[
\sup_{t\geq0}|J_{N,k,l}^{2}(t)|\leq\mathrm{const}\cdot N\cdot\sum_{m=[N/l]}^{2N-1}\frac{1}{m^{2}}\leq\mathrm{const}\cdot N\int_{N/l}^{\infty}\frac{\mathrm{d}x}{x^{2}}=\mathrm{const}\cdot l\leq\mathrm{const}\cdot F_{N}(l)
\]
}

In the first inequality we used the assertion 3 of lemma 3, in the
second we used ($\ref{eq:fact2}$), and in the last one we used the
definition ($\ref{eq:F}$). Joining the results together we have the
desired bound from above 
\[
\sup_{t\geq0}|I_{N,k,l}(t)|\leq\sup_{t\geq0}|J_{N,k,l}^{1}(t)|+\sup_{t\geq0}|J_{N,k,l}^{2}(t)|\leq\mathrm{const}\cdot F_{N}(l).
\]

\subsection{Proof of lemma 1}

We have three equations for the potential at the point $a=\frac{1}{N}$
\begin{equation}
V(a)=-c_{n}N^{n}+c_{m}N^{m}\label{a0}
\end{equation}
\begin{equation}
V'(a)=0=nc_{n}N^{n+!}-mc_{m}N^{m+1}\label{a1}
\end{equation}
\begin{equation}
V''(a)=\kappa=-n(n+1)c_{n}N^{n+2}+m(m+1)c_{m}N^{m+2}\label{a2}
\end{equation}
However, in the first equation we do not know the value $V(a)$. From
(\ref{a0}) and (\ref{a1}) we have
\begin{equation}
c_{m}=\frac{n}{m-n}N^{-m}V(a),c_{n}=\frac{1}{m-n}N^{-n}V(a)\label{c_U}
\end{equation}
and from (\ref{a1}) and (\ref{a2})
\begin{equation}
c_{m}=\frac{\kappa}{m(m-n)}N^{-2-m},c_{n}=\frac{\kappa}{n(m-n)}N^{-2-n}\label{c_kappa}
\end{equation}
From this we can find $V(a)=-\frac{\kappa}{mn}N^{-2}$. Also we need
the inflection point $b>a>0$ , which can be found from the condition
\begin{equation}
V''(b)=-n(n+1)c_{n}b^{-(n+2)}+m(m+1)c_{m}b^{-(m+2)}=0,\label{U_2_b}
\end{equation}
thus
\[
b=\frac{1}{N}(\frac{m+1}{n+1})^{\frac{1}{m-n}}
\]
and
\[
V'(b)=nc_{n}b^{-n-1}-mc_{m}b^{-m-1}
\]
Using (\ref{a1}) together with the equation (\ref{fixed_point})
for $h=b$, we find $\sigma(N)=\frac{f}{\kappa}$. From (\ref{U_2_b})
we have
\[
\frac{f}{\kappa}=\frac{C}{N},C=C(n,m)=\frac{1}{m-n}((\frac{m+1}{n+1})^{-\frac{n-1}{m-n}}-(\frac{m+1}{n+1})^{-\frac{m-1}{m-n}})
\]

\end{document}